\documentclass[aps,prl,reprint,preprintnumbers,superscriptaddress,amsmath,amssymb,bibnotes,longbibliography]{revtex4-2}
\usepackage{graphicx}
\usepackage{dcolumn}
\usepackage{bm}
\usepackage[colorlinks,linkcolor=blue,anchorcolor=blue,citecolor=blue,breaklinks,CJKbookmarks=True,urlcolor=blue,filecolor=blue,menucolor=blue,runcolor=blue]{hyperref}

\begin{document}

\title{A Family of Lanthanide Noncentrosymmetric Superconductors La$_4$$TX$ ($T$ = Ru, Rh, Ir; $X$ = Al, In)}

\author{Hang Su}
\affiliation  {Center for Correlated Matter and Department of Physics, Zhejiang University, Hangzhou 310058, China}
\affiliation  {Zhejiang Province Key Laboratory of Quantum Technology and Device, Department of Physics, Zhejiang University, Hangzhou 310058, China}
\author{Feng Du}
\affiliation  {Center for Correlated Matter and Department of Physics, Zhejiang University, Hangzhou 310058, China}
\affiliation  {Zhejiang Province Key Laboratory of Quantum Technology and Device, Department of Physics, Zhejiang University, Hangzhou 310058, China}
\author{Shuaishuai Luo}
\affiliation  {Center for Correlated Matter and Department of Physics, Zhejiang University, Hangzhou 310058, China}
\affiliation  {Zhejiang Province Key Laboratory of Quantum Technology and Device, Department of Physics, Zhejiang University, Hangzhou 310058, China}
\author{Zhiyong Nie}
\affiliation  {Center for Correlated Matter and Department of Physics, Zhejiang University, Hangzhou 310058, China}
\affiliation  {Zhejiang Province Key Laboratory of Quantum Technology and Device, Department of Physics, Zhejiang University, Hangzhou 310058, China}
\author{Rui Li}
\affiliation  {Center for Correlated Matter and Department of Physics, Zhejiang University, Hangzhou 310058, China}
\affiliation  {Zhejiang Province Key Laboratory of Quantum Technology and Device, Department of Physics, Zhejiang University, Hangzhou 310058, China}
\author{Wu Xie}
\affiliation  {Center for Correlated Matter and Department of Physics, Zhejiang University, Hangzhou 310058, China}
\affiliation  {Zhejiang Province Key Laboratory of Quantum Technology and Device, Department of Physics, Zhejiang University, Hangzhou 310058, China}
\author{Bin Shen}
\affiliation  {Center for Correlated Matter and Department of Physics, Zhejiang University, Hangzhou 310058, China}
\affiliation  {Zhejiang Province Key Laboratory of Quantum Technology and Device, Department of Physics, Zhejiang University, Hangzhou 310058, China}
\author{Yunfeng Wang}
\affiliation  {Center for Correlated Matter and Department of Physics, Zhejiang University, Hangzhou 310058, China}
\author{An Wang}
\affiliation  {Center for Correlated Matter and Department of Physics, Zhejiang University, Hangzhou 310058, China}
\affiliation  {Zhejiang Province Key Laboratory of Quantum Technology and Device, Department of Physics, Zhejiang University, Hangzhou 310058, China}
\author{Toshiro Takabatake}
\affiliation  {Center for Correlated Matter and Department of Physics, Zhejiang University, Hangzhou 310058, China}
\affiliation  {Department of Quantum Matter, Graduate School of Advanced Science and Engineering, Hiroshima University, Higashi-Hiroshima 739-8530, Japan}
\author{Chao Cao}
\affiliation  {Center for Correlated Matter and Department of Physics, Zhejiang University, Hangzhou 310058, China}
\affiliation  {Condensed Matter Group, Department of Physics, Hangzhou Normal University, Hangzhou 311121, China}
\author{Michael Smidman}
\email[Corresponding author: ]{msmidman@zju.edu.cn}
\affiliation  {Center for Correlated Matter and Department of Physics, Zhejiang University, Hangzhou 310058, China}
\affiliation  {Zhejiang Province Key Laboratory of Quantum Technology and Device, Department of Physics, Zhejiang University, Hangzhou 310058, China}
\author{Huiqiu Yuan}
\email[Corresponding author: ]{hqyuan@zju.edu.cn}
\affiliation  {Center for Correlated Matter and Department of Physics, Zhejiang University, Hangzhou 310058, China}
\affiliation  {Zhejiang Province Key Laboratory of Quantum Technology and Device, Department of Physics, Zhejiang University, Hangzhou 310058, China}
\affiliation  {State Key Laboratory of Silicon Materials, Zhejiang University, Hangzhou 310058, China}
\affiliation  {Collaborative Innovation Center of Advanced Microstructures, Nanjing University, Nanjing, 210093, China}

\date{\today}

\begin{abstract}
We report the discovery of superconductivity in a series of  noncentrosymmetric compounds La$_4$$TX$ ($T$ = Ru, Rh, Ir; $X$ = Al, In), which have a cubic crystal structure with space group $F\bar{4}3m$. La$_4$RuAl, La$_4$RhAl, La$_4$IrAl, La$_4$RuIn and La$_4$IrIn exhibit bulk superconducting transitions with critical temperatures $T_c$ of  1.77~K, 3.05~K, 1.54~K, 0.58~K and 0.93~K, respectively. The specific heat of the La$_4$$T$Al compounds are consistent with an $s$-wave model with a fully open superconducting gap. In all cases, the upper critical fields are well described by the Werthamer-Helfand-Hohenberg model, and the values are well below the Pauli limit, indicating that orbital limiting is the dominant pair-breaking mechanism. Density functional theory (DFT) calculations reveal that the degree of band splitting by the  antisymmetric spin-orbit coupling (ASOC) shows considerable variation between the different compounds. This indicates that the strength  of the ASOC is highly tunable across this series of  superconductors, suggesting that these are good candidates for examining the relationship between the ASOC and superconducting properties in noncentrosymmetric superconductors.

\end{abstract}

\maketitle

\section{\uppercase\expandafter{\romannumeral1}. INTRODUCTION}

Noncentrosymmetric superconductors (NCS), where the crystal structure lacks an inversion center, have been identified as prime candidates for realizing unconventional superconducting pairing states \cite{Smidman2017}, following the discovery of unconventional superconductivity in the heavy fermion antiferromagnet CePt$_3$Si \cite{BauerPhysRevLett2003}. When inversion symmetry is broken, an antisymmetric spin-orbit coupling (ASOC) can lift the degeneracy of the electrons near the Fermi level, allowing for superconductivity with a mixture of spin-singlet and spin-triplet pairing \cite{GorkovPhysRevLett2001}. Subsequently, pressure induced superconductivity was realized in the noncentrosymmetric heavy fermion systems Ce(Rh, Ir)Si$_3$ \cite{KimuraPhysRevLett2005, sugitani2006pressure}, Ce(Co, Ir, Rh)Ge$_3$ \cite{SETTAI2007844, HONDA2010S543, WangPhysRevB2018}, and UIr \cite{Akazawa2004}, where large and anisotropic upper critical fields have been observed well in excess of the Pauli limit \cite{KimuraPhysRevLett2007, SettaiJPSJ2008, MeassonJPSJ2009}, but disentangling the influence on the superconductivity of the inversion symmetry breaking from the strong electronic correlations and magnetism has proved challenging. As such, weakly correlated noncentrosymmetric superconductors have also been investigated. For instance, in the Li(Pd$_{1-x}$Pt$_x$)$_3$B system, Li$_2$Pd$_3$B has two nodeless superconducting gaps, while in Li$_2$Pt$_3$B there is evidence for one of the gaps having line nodes, which has been ascribed to stronger singlet-triplet mixing upon increasing the ASOC, by replacing Pd with the heavier element Pt \cite{YuanPhysRevLett2006}. On the other hand, the properties of many reported weakly correlated NCS's are consistent with purely $s$-wave pairing, including La(Rh, Pt, Pd, Ir)Si$_3$ \cite{AnandPhysRevB2011, SmidmanPhysRevB2014, OkudaJPSJ2007, AnandPhysRevB2014}, (Rh, Ir)$_2$Ga$_9$ \cite{HirokoJPSJ2007, KouheiJPSJ2009}, Re-$T$ ($T$ = transition metal) \cite{SinghPhysRevLett2014, SinghPhysRevB2018, SinghDPhysRevB2016, ShangPhysRevBReTi2018, ChenJPhysRevB2013}, and BiPd \cite{JoshiPhysRevB2011, Jiao2014}, despite a number of these systems exhibiting an appreciable band splitting due to the ASOC \cite{Smidman2017}. As such, the relationship between the ASOC and nature of the superconducting order parameter, as well as the necessary conditions for significant singlet-triplet mixing, still remains to be determined, requiring the study of additional classes of NCS where the ASOC can be tuned.

In recent years, time-reversal symmetry (TRS) breaking has also been detected in the superconducting states of a number of NCS \cite{Ghosh2020}, including LaNiC$_2$ \cite{HillierPhysRevLett2009}, La$_7$(Ir, Rh)$_3$ \cite{SinghDPhysRevB2020, BarkerPhysRevLett2015}, Re-$T$ \cite{SinghPhysRevLett2014, SinghPhysRevB2017, SinghPhysRevB2018, ShangPhysRevLett2018} and CaPtAs \cite{Xie2020, ShangPhysRevLett2020}, but in most cases the relationship between the broken time reversal and inversion symmetries is not resolved.
In systems such as orthorhombic LaNiC$_2$, the low symmetry of the crystal structure excludes there being both TRSB and significant singlet triplet mixing \cite{QuintanillaPhysRevB2010}, which led to the proposal of an internally-antisymmetric, non-unitary triplet state \cite{WengPhysRevLett2017}. On the other hand, in higher symmetry structures such as  the cubic $\alpha$-Mn type structure of several Re-based NCS, there are symmetry allowed mixed parity states with TRSB \cite{SinghPhysRevLett2014}, but the actual origin of the TRSB in these materials is yet to be determined. As such, it is of considerable interest to search for more families of NCS, in particular those crystallizing in cubic structures.

\begin{figure}
  	\includegraphics[angle=0,width=0.49\textwidth]{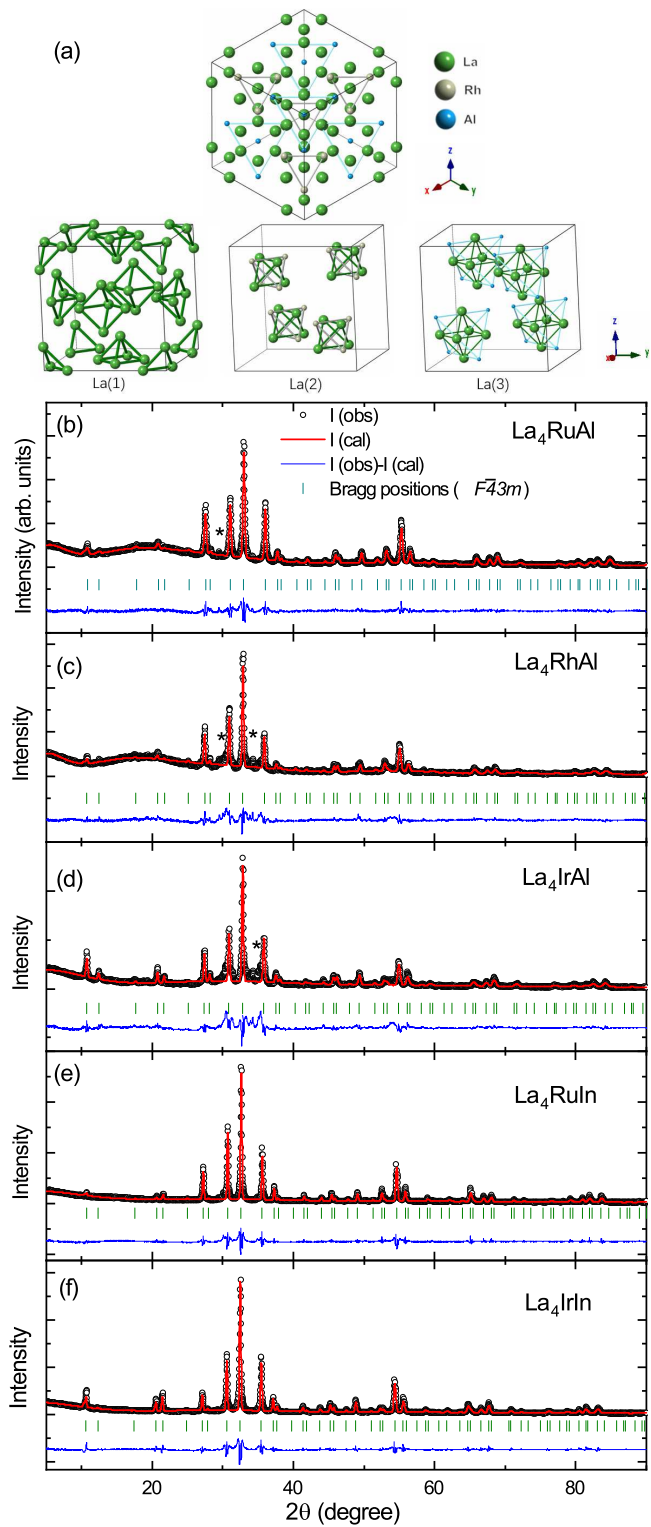}
  	\vspace{-12pt} \caption{\label{Figure1} (a) Crystal structure of La$_4$RhAl. Atoms occupying different crystallographic sites are also shown separately. Powder x-ray diffraction patterns of polycrystalline (b) La$_4$RuAl (c) La$_4$RhAl (d) La$_4$IrAl (e) La$_4$RuIn and (f) La$_4$IrIn. The red solid lines denote calculated results from Rietveld refinements using the Gd$_4$RhIn-type structure and the blue solid lines denote the difference between the observed and calculated patterns. Small unindexed peaks are marked by an asterisk. The calculated Bragg peak positions are shown by the green vertical lines. }
  \end{figure}

The La$_4TX$ ($T$ = Ru, Rh, Ir; $X$ = Al, In) compounds belong to a large family of rare-earth based materials \cite{tappe2011new}, which have the cubic Gd$_4$RhIn-type structure with the noncentrosymmetric space group $F\bar{4}3m$ \cite{zaremba2007rare}. The Gd$_4$RhIn-type  structure is a ternary ordered variant of the centrosymmetric Ti$_2$Ni-type structure (space group $Fd\bar{3}m$) \cite{solokha2009crystal, POTTGEN20201}. The  crystal structure of La$_4TX$ is displayed in Fig. \ref{Figure1}(a) \cite{tappe2011new}. Slightly distorted La$_6T$ trigonal prisms form a rigid three dimensional network sharing corners and edges, with cavities that are occupied by La octahedra and $X$ tetrahedra, where the $T$ and $X$ atoms are located on noncentrosymmetric sites. There are three inequivalent La sites, among which only the La(1) are centrosymmetric. Superconductivity has been reported in systems with a different structure but the same space group as the Gd$_4$RhIn-type, namely in the equiatomic half-Heusler compounds including nonmagnetic (Y, Lu)(Pd, Pt)Bi with topological electronic structures \cite{wang2013large, xu2014weak, ButchPhysRevB2011, TaftiPhysRevB2013} and the magnetic superconductors $R$PdBi ($R$: magnetic rare earth) \cite{pan2013superconductivity, Nakajimae1500242}, among which evidence for unconventional pairing states are also reported \cite{MeinertPhysRevLett2016, RadmaneshPhysRevB2018, Kimeaao4513}.

In this study, we report the discovery of superconductivity in five La$_4TX$ compounds using resistivity, magnetization and specific heat measurements, as well as electronic structure calculations. The occurrence of superconductivity in several systems with different atoms on the $T$ and $X$ sites suggest that the La$_4TX$ superconductors are good candidates for examining the effects of tuning the ASOC on the superconducting and normal state properties.

\section{\uppercase\expandafter{\romannumeral2}. EXPERIMENTAL METHODS}

Polycrystalline samples of La$_4$$TX$ ($T$ = Ru, Rh, Ir; $X$ = Ru, In.) were synthesized by arc-melting stoichiometric amounts of the constituent elements. In order to improve the sample quality, La and $T$ were first melted together and the resulting boule was subsequently melted with $X$. After being turned over and remelted several times to ensure homogeneity, the samples were sealed in evacuated quartz ampoules and annealed for three weeks at 350~$^{\circ}$C and 700~$^{\circ}$C for $X$ = Al and In, respectively, before being quenched in water. Note that the attempted synthesis of La$_4$RhIn by the above method was unsuccessful, where only a mixture of binary compounds was obtained. The crystal structures at room temperature were confirmed by powder x-ray diffraction using a Rigaku Ultima IV diffractometer with Cu K$\alpha$ radiation. Resistivity measurements were performed using a standard four-probe method on a Physical Property Measurement System (PPMS, Quantum Design) with $^3$He option as well as an Oxford Instruments $^3$He refrigerator. The specific heat was also measured down to 0.4~K using the relaxation method in a PPMS. The ac magnetic susceptibility was measured down to 0.3~K using the mutual-inductance method with an ac magnetometer in a $^3$He refrigerator (Oxford Instruments), to characterize the superconducting transitions. DC magnetization measurements were carried out using a superconducting quantum interference device (SQUID) magnetometer (MPMS) down to 2~K. Density functional theory (DFT) calculations were performed using the Vienna ab-initio simulation package (VASP) to obtain the electronic band structure and density of states. The Perdew-Burke-Ernzerhoff (PBE) functional in the generalized gradient approximation (GGA) was applied. The Brillouin zone was sampled with $12\times12\times12$ K meshes and the energy-cutoff was 450~eV.

\begin{table}[tb]
\caption{Results of the Rietveld refinements of the powder x-ray diffraction results. The lattice parameters, profile factors, and atomic positions are shown.}
\label{table:table1}
\begin{ruledtabular}
 \begin{tabular}{c c  c c }
{} &$a$    &$R_p$  &$R_{wp}$   \\
{} &(\AA)        &($\%$)     &($\%$)                                      \\
\hline\\[-2ex]
La$_4$RuAl     &{14.0814(2)}            &7.98    &10.72        \\
La$_4$RhAl    &{14.1826(4)}             &10.43     &14.41           \\
La$_4$IrAl    &{14.1741(5)}             &16.26     &22.42       \\
La$_4$RuIn    &{14.2428(2)}             &12.88   &17.26    \\
La$_4$IrIn     &{14.3532(2)}            &13.23     &18.31       \\
{}              &{}                      &{}       &{}          \\
{}              &{x}                     &{y}      &{z}          \\
\hline\\[-2ex]
{La(1)}              &{0.5579(3)}             &{0.25}    &{0.25}   \\
{La(2)}             &{0.1924(3)}             &{0}        &{0}     \\
{La(3)}             &{0.3509(2)}        &{0.3509(2)}    &{0.3509(2)}     \\
{Ru}             &{0.1375(2)}          &{0.1375(2)}   &{0.1375(2)}     \\
{Al}             &{0.5702(7)}           &{0.5702(7)}  &{0.5702(7)}    \\
\end{tabular}
\end{ruledtabular}
\end{table}

  \begin{figure*}
  	\includegraphics[angle=0,width=0.98\textwidth]{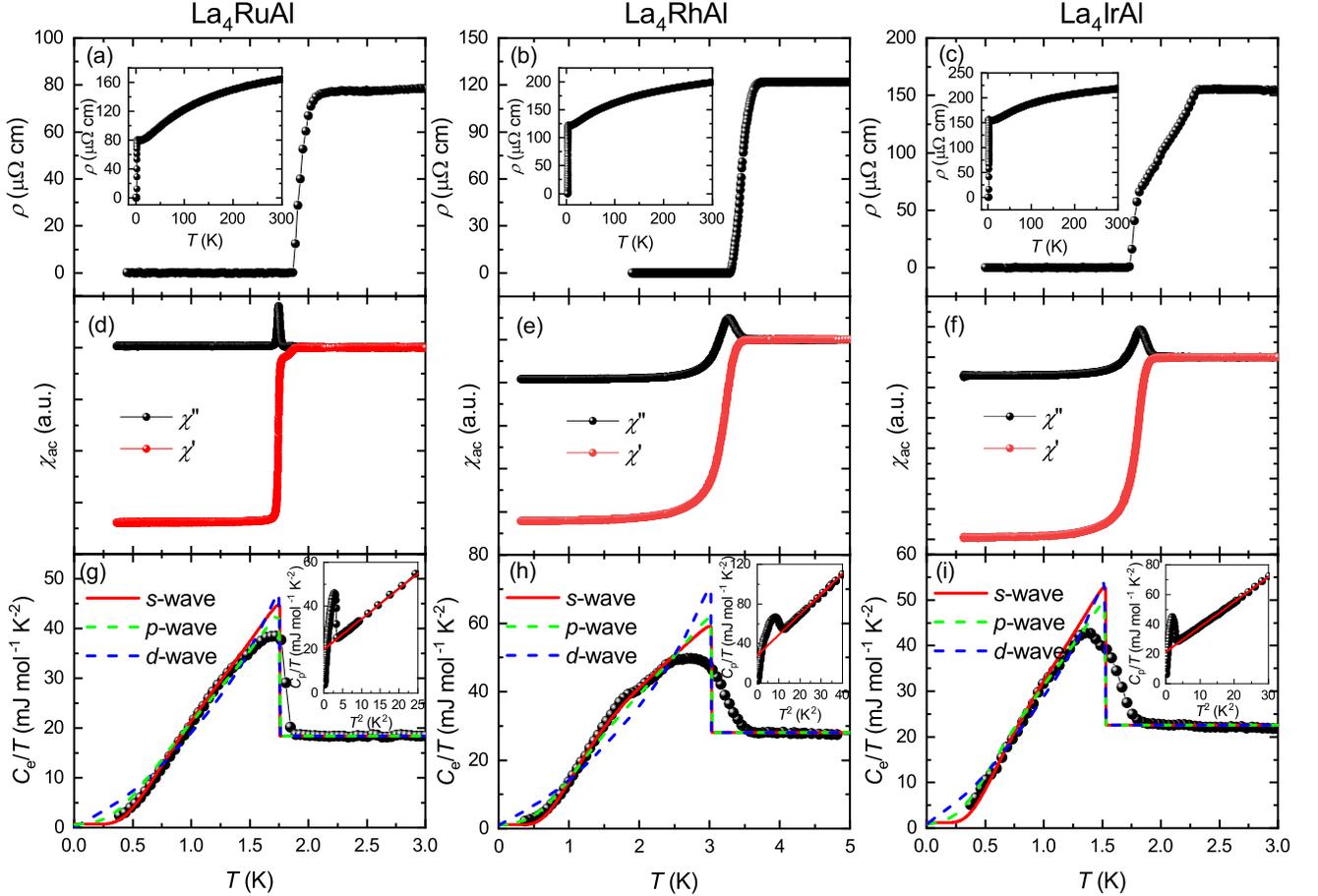}
  	\vspace{-12pt} \caption{\label{Figure2}  Characterization of the superconducting transitions of La$_4$RuAl, La$_4$RhAl, and La$_4$IrAl in zero field. (a) (b) and (c) show the respective temperature dependences of the resistivity near the transition, while the insets show the data up to room temperature. (d) (e) and (f) show the corresponding real and imaginary parts of the ac susceptibility, while (g) (h) and (i) display the electronic specific heat after subtracting the phonon contribution. The red solid lines and green and blue dashed lines show the results from fitting, respectively, with $s$-wave, $p$-wave, and $d$-wave models described in the text. The insets display $C_p/T$ versus $T^2$ at low temperatures, where the red solid lines show the results from fitting to $C_p/T = \gamma_n+\beta T^2$.}
  	\vspace{-12pt}
  \end{figure*}

  \begin{figure}
  	\includegraphics[angle=0,width=0.49\textwidth]{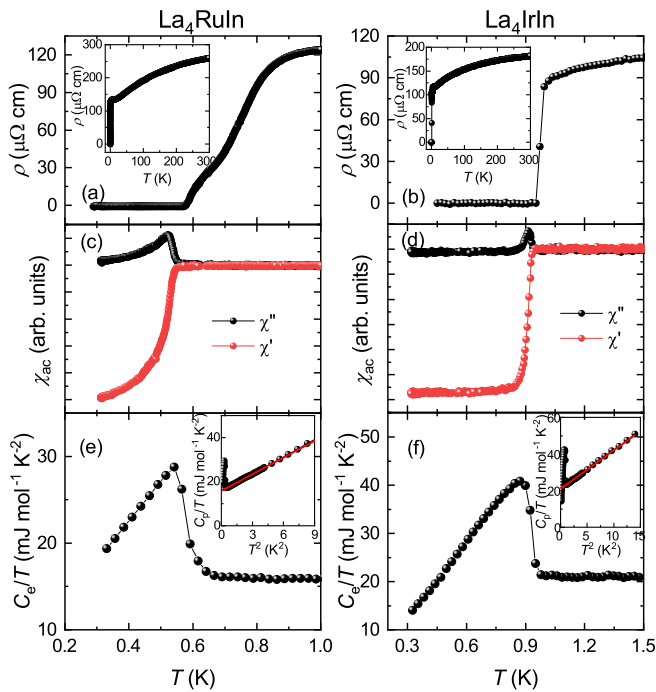}
  	\vspace{-12pt} \caption{\label{Figure3}  Temperature dependence of the resistivity, ac susceptibility and electronic specific heat of La$_4$RuIn and La$_4$IrIn near $T_c$. Insets of (a) and (b): resistivity versus temperature from 0.5~K to 300~K. Insets of (e) and (f): Low temperature $C_p/T$ versus $T^2$, the red solid lines show the fits to the data with the form $C_p/T = \gamma_n+\beta T^2$.}
  	\vspace{-12pt}
  \end{figure}

  \begin{figure*}
  	\includegraphics[angle=0,width=0.98\textwidth]{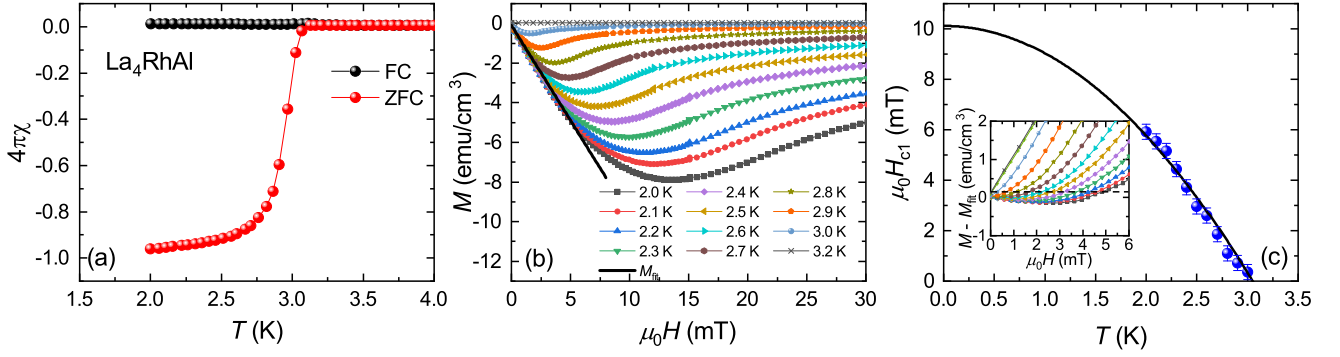}
  	\vspace{-12pt} \caption{\label{Figure4} (a) Zero-field-cooled (ZFC) and field-cooled (FC) temperature dependence of the magnetic susceptibility of La$_4$RhAl measured in an applied magnetic field of 1~mT. (b) Field dependence of the magnetization at various temperatures below $T_c$. The black solid line shows the linear fitting in the low field region. (c) Lower critical field $\mu_0H_{c1}$ versus temperature of La$_4$RhAl. The solid line is the fitting result using  Eq. (\ref{equation5}). The inset illustrates the determination of $\mu_0H_{c1}$ (see text).}
  	\vspace{-12pt}
  \end{figure*}

  \begin{figure*}
  	\includegraphics[angle=0,width=0.98\textwidth]{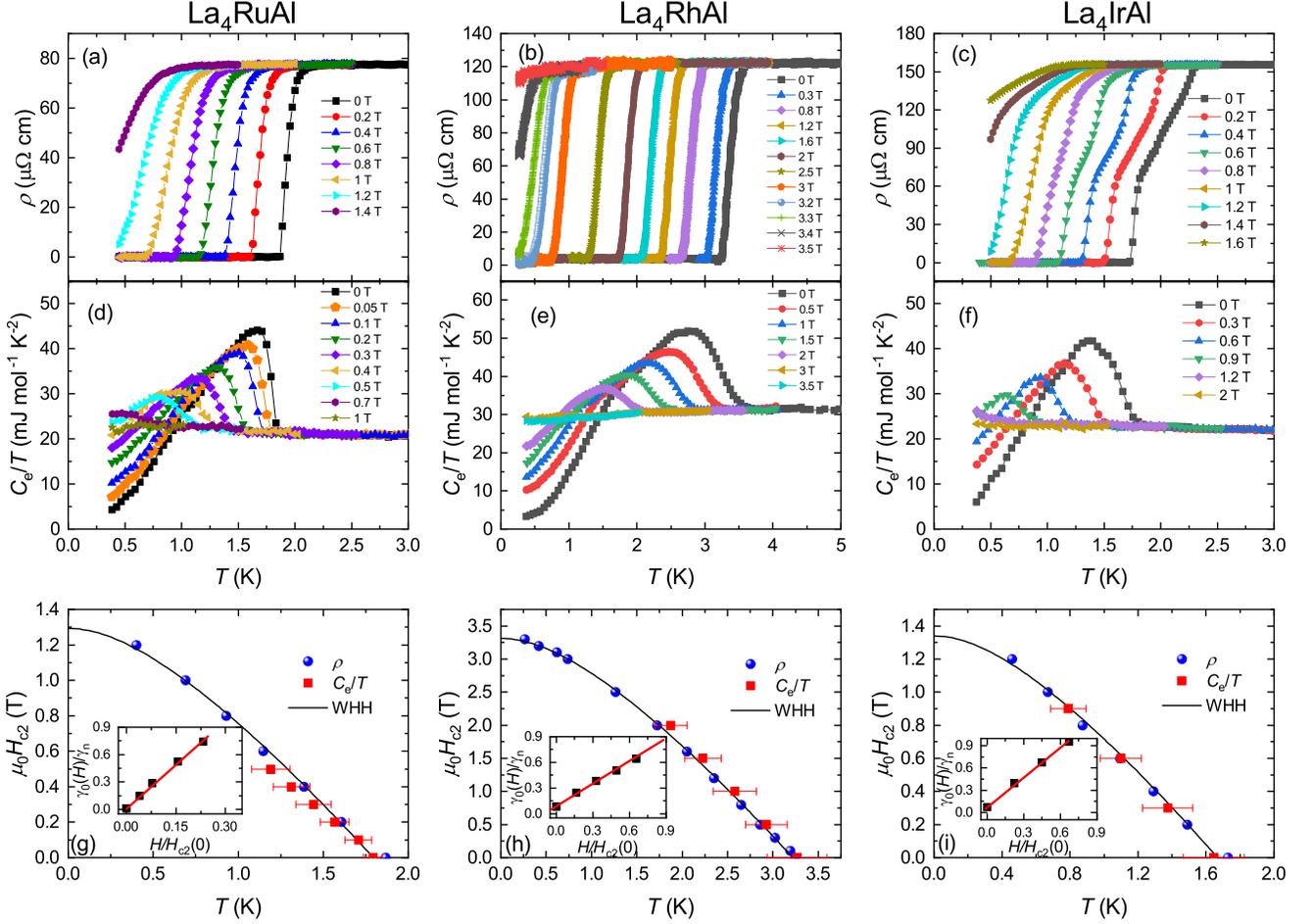}
  	\vspace{-12pt} \caption{\label{Figure5}  Temperature dependence of (a), (b), (c): resistivity and (d), (e), (f): electronic specific heat of La$_4$RuAl, La$_4$RhAl, and La$_4$IrAl, respectively, measured under various applied magnetic fields in the vicinity of $T_c$. The upper critical field $\mu_0H_{c2}$ versus $T$ are respectively shown in (g), (h), and (i). $T_c$ was determined from where there is zero resistivity and the midpoint of the specific heat transition. The solid lines correspond to fitting using the WHH model. The insets of panel (g)-(i) show the field dependence of the residual Sommerfeld coefficient $\gamma_0$, plotted as $\gamma_0(H)/\gamma_n$ versus $H/H_{c2}(0)$, where the values of $\gamma_0(H)$ are estimated by extrapolating $C/T \sim T^2$ curves to zero temperature.}
  	\vspace{-12pt}
  \end{figure*}

  \begin{figure}
  	\includegraphics[angle=0,width=0.49\textwidth]{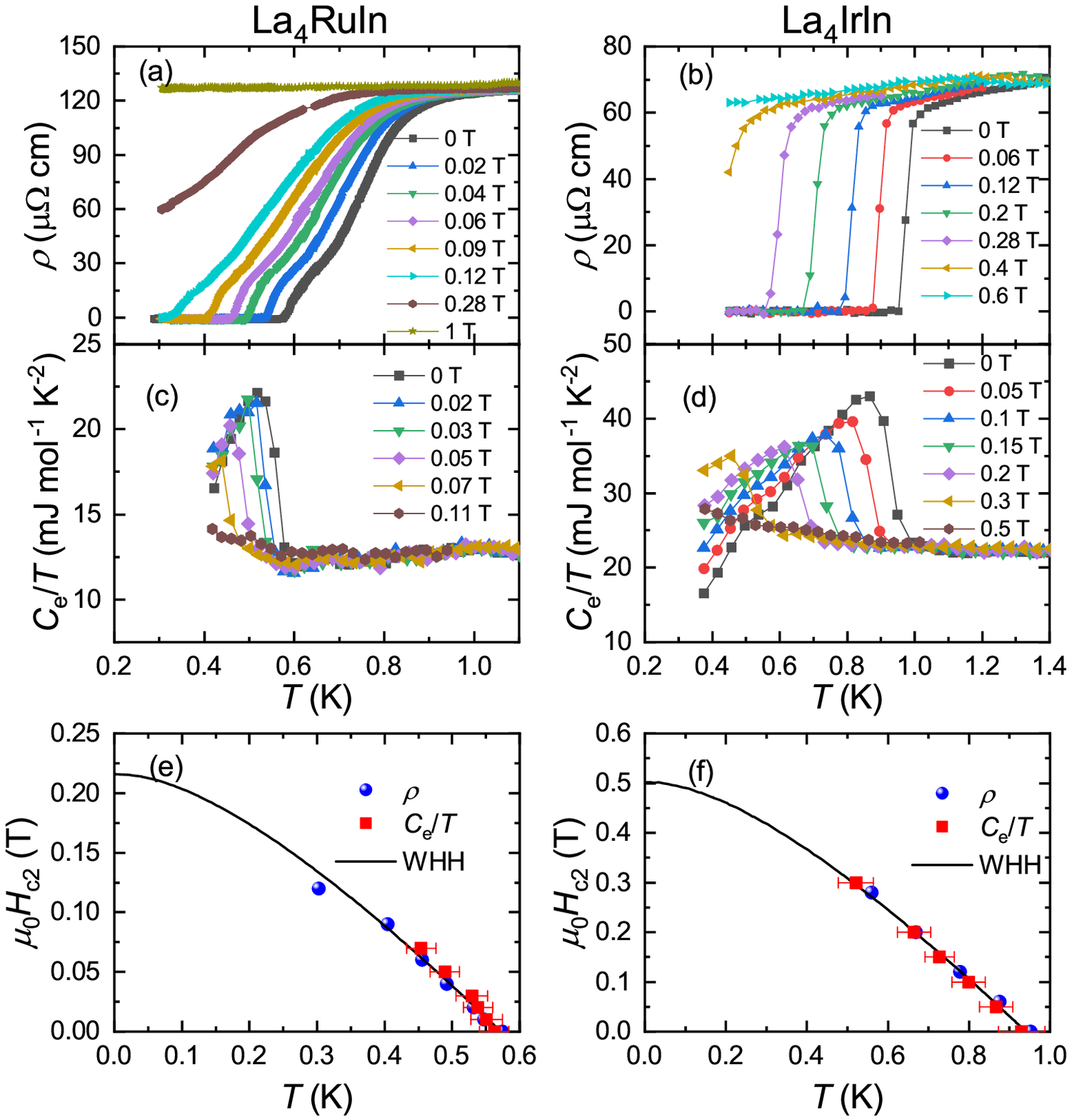}
  	\vspace{-12pt} \caption{\label{Figure6}  Temperature dependence of the resistivity and specific heat of La$_4$RuIn and La$_4$IrIn in various applied magnetic fields. (e) and (f) show the corresponding upper critical fields, where $T_c$ was determined from the onset of zero resistivity and the midpoints of the specific heat jump. The solid lines are the fits using the WHH model.}
  	\vspace{-12pt}
  \end{figure}

\section{\uppercase\expandafter{\romannumeral3}. Results and Discussion}

\subsection{\uppercase\expandafter{a}. Crystal structure}

X-ray diffraction patterns of La$_4$$T$Al ($T$ = Ru, Rh, Ir) and La$_4$$T$In ($T$ = Ru, In) are shown in Figs. \ref{Figure1}(b)-(f). All the experimental patterns are well indexed by the noncentrosymmetric Gd$_4$RhIn-type cubic structure \cite{zaremba2007rare},
and they can be well-refined using this structural model. The results are displayed in Table. \ref{table:table1}, where the fitted lattice parameters are in good agreement with the reported values \cite{tappe2011new}. The atomic positions for La$_4$RuAl are also displayed in the table, and very similar values are obtained for the other four compounds.
We note that although the Gd$_4$RhIn-type structure is an ordered variant of the Ti$_2$Ni-type structure, the latter leads to very different peak intensities, and therefore our XRD data cannot be refined with a Ti$_2$Ni-type  model.
The small additional peaks in the La$_4$$T$Al patterns may correspond to the binary phases La$_5$Al$_4$ or La$_{16}$Al$_{13}$, likely as a result of the incongruent melting of the ternary phases.

\subsection{\uppercase\expandafter{b}. Zero-field superconducting properties }

The temperature dependence of the electrical resistivity of La$_4$$T$Al ($T$ = Ru, Rh and Ir) are shown in Figs. \ref{Figure2}(a)-(c), in the vicinity of the superconducting transition temperatures $T_c$. The insets show the resistivity across the whole temperature range, and all the samples exhibit metallic behavior. The relatively large residual resistivity in the normal state $\rho_0$ (77$-$155~$\mu\Omega$~cm) and the small residual resistivity ratio RRR = $\rho$(300~K)/$\rho$(2~K) = 1.4$-$2.0 are possibly due to atomic disorder on the nominal Al $16e$ site \cite{tappe2011new}. At low temperatures, the resistivity drops to zero at 1.86~K, 3.28~K, and 1.73~K for La$_4$RuAl, La$_4$RhAl, and La$_4$IrAl, respectively. For La$_4$IrAl, the transition width is relatively broad, and there is a kink in the resistivity data, which could be due to sample inhomogeneity. The temperature dependence of the ac susceptibility are shown in Figs. \ref{Figure2}(d)-(f). An abrupt decrease of the real part and a peak in the imaginary part indicate the presence of a superconducting transition, with $T_c$ values similar to those determined from the resistivity.

The low temperature electronic specific heat $C_e/T$ for each compound is displayed in Figs. \ref{Figure2}(g)-(i), after subtracting the phonon contribution $\beta T^2$, which was determined from fitting the total specific heat in the normal state with $C_p/T = \gamma_n+\beta T^2$, where $\gamma_n$ denotes the Sommerfeld coefficient. The fitting yields $\gamma_n$ = 20.1(1), 29.3(2), and 21.3(1)~mJ~mol$^{-1}$~K$^{-2}$, $\beta$ = 1.35(3), 1.29(3) and 1.68(2)~mJ~mol$^{-1}$~K$^{-4}$ for La$_4$RuAl, La$_4$RhAl, and La$_4$IrAl, respectively. The relatively low values of $\gamma_n$ suggest weak electronic correlations. The Debye temperatures $\theta_D$ are estimated from $\beta$ using $\theta_D = (12\pi^4Rn/5\beta)^{1/3}$, with the number of atoms per formula unit $n$ = 6 and the molar gas constant $R$ = 8.314~J~mol$^{-1}$~K$^{-1}$, yielding 204(2)~K, 208(2)~K and 191(1)~K for the respective materials. The presence of transitions in the electronic specific heat indicate bulk superconductivity, with $T_c$ values close to those found in the resistivity and ac susceptibility. The normalized specific heat jumps at $T_c$, $\Delta C/\gamma_nT_c$ are 1.23, 0.93 and 1.09 for the respective Ru, Rh and Ir variants, which are smaller than the BCS weak coupling limit value of 1.43. The deviation may be ascribed to the broad nature of the transitions, as well as reduced or anisotropic gap magnitudes. The electron-phonon coupling constant $\lambda_{\mathrm{el-ph}}$ can be estimated from McMillan's theory \cite{McMillanPhysRev1968} by
\begin{equation}\label{equation1}
\lambda_{\mathrm{el-ph}}=\frac{1.04+\mu^*\mathrm{ln}(\theta_D/1.45T_c)}{(1-0.62\mu^*)\mathrm{ln}(\theta_D/1.45T_c)-1.04}
\end{equation}
yielding 0.45$-$0.54, 0.52$-$0.62 and 0.44$-$0.53, for a screened Coulomb parameter of $\mu$ = 0.1$-$0.15. The relatively small values of $\lambda_{\mathrm{el-ph}}$ suggest weak electron-phonon coupling. Among the three samples, a comparatively larger value of $\lambda_{\mathrm{el-ph}}$ is correlated with a higher value of $T_c$.

The temperature dependence of the electronic specific heat can be expressed as $C_e$/$T$=d$S$/d$T$, where $S(T)$ is the superconducting contribution to the entropy.
$S(T)$ can be calculated using \cite{tinkham2004introduction}
\begin{equation}\label{equation2}
S(T)=-\frac{3\gamma_n}{k_B\pi^3}\int_{0}^{2\pi}\int_{0}^{\infty}[(1-f)\mathrm{ln}(1-f)+f\mathrm{ln}f]\mathrm{d}\epsilon\mathrm{d}\phi
\end{equation}
\begin{equation}\label{equation3}
f = [1+\mathrm{exp}(\sqrt{\epsilon^2+\Delta^2(T)}/k_BT)]^{-1}
\end{equation}
\begin{equation}\label{equation4}
\Delta(T) = g_k\Delta(0)\tanh{1.82[1.018((T_c/T)-1)]^{0.51}}
\end{equation}
where $f$ is the Fermi function, $\epsilon$ is the relative electronic energy, and $\Delta(0)$ is the superconducting gap at zero temperature. For the angle-dependent component $g_k$, values of 1, sin$\theta$, and cos$2\phi$ are used for the $s$-wave, $p$-wave, and $d$-wave models, respectively, where $\theta$ is the polar angle and $\phi$ is the azimuthal angle. The experimental data for all three compounds are well fitted by the weak-coupling $s$-wave model shown by the red lines in Fig. \ref{Figure2}, with $\Delta(0)$ = 1.78$k_BT_c$ for La$_4$RuAl, 1.57$k_BT_c$ for La$_4$RhAl, and 1.74$k_BT_c$ for La$_4$IrAl. The data are not well described by the $p$-wave and $d$-wave models (dashed lines), and therefore the data are more consistent with a nodeless $s$-wave model, rather than the two models with a nodal gap. On the other hand, in order to confirm fully gapped behavior and precisely constrain the gap magnitude, further measurements to lower temperatures are necessary, especially in the case of La$_4$IrAl which has the lowest $T_c$.

We also investigated the superconducting properties of the $X$ = In compounds, La$_4$RuIn and La$_4$IrIn. The temperature dependences of the resistivity, ac susceptibility and electronic specific heat are shown in Fig. \ref{Figure3}. Both exhibit metallic behavior, as shown in the insets, with respective residual resistivities of 134 and 100~$\mu\Omega$~cm, and RRR of 1.9 and 1.6. La$_4$RuIn has a superconducting transition at around 0.58~K, with a relatively wide transition width, while La$_4$IrIn becomes superconducting below around 0.93~K. As shown in the insets, fitting the low temperature specific heat above $T_c$ yields $\gamma_n$ of 15.9(1) and 21.1(2)~mJ~mol$^{-1}$~K$^{-2}$ for La$_4$RuIn and La$_4$IrIn, together with $\beta$ of 2.51(2) and 2.09(2)~mJ~mol$^{-1}$~K$^{-4}$. These correspond to $\theta_D$ of 169(1) and 177(1)~K, which are lower than those of the $X$ = Al compounds described above, consistent with the lower values of $T_c$. Using Eq. (\ref{equation1}), $\lambda_{\mathrm{el-ph}}$ were estimated to be 0.36$-$0.45 and 0.39$-$0.48, implying weaker electron-phonon coupling strengths compared with the $X$ = Al compounds. The electronic specific heat jumps $\Delta C/\gamma_nT_c$ of 0.82 for La$_4$RuIn and 1.01 for La$_4$IrIn are much smaller than the BCS value of 1.43. Polycrystalline samples of isostructural Lu$_4$RhAl and Lu$_4$RhIn were also successfully synthesized, and no superconductivity was observed from measurements down to 0.4~K.

\subsection{\uppercase\expandafter{c}. In-field superconducting properties}

Measurements of the dc magnetic susceptibility were carried out only for La$_4$RhAl as its $T_c$ is higher than the low-temperature limit (2~K) of the available MPMS. Figure. \ref{Figure4}(a) displays the low temperature dc magnetic susceptibility of La$_4$RhAl upon both zero-field cooling (ZFC) and field-cooling (FC) in an applied field of 1~mT, where the data are corrected for demagnetization effects. Here a demagnetization factor of $N=0.18$ is estimated, considering a cuboid sample of dimensions $0.81\times0.55\times1.45$~ mm$^3$, with the field applied along the longest side \cite{Aharoni1998}. A clear transition at around 3.1~K and a saturated diamagnetic signal value of $4\pi\chi$ close to $-1$ in the ZFC curve provide evidence for full diamagnetic shielding in La$_4$RhAl. The large difference between the FC and ZFC curves is due to trapping of the magnetic flux, which is frequently observed in polycrystalline samples of type-II superconductors. Figure. \ref{Figure4}(b) displays the field dependence of the magnetization $M(H)$ measured at various temperatures below $T_c$ using a ZFC protocol, which shows the typical behavior expected for a type-II superconductor. A linear fitting of the low-field region, $M_{\mathrm{fit}}$, is shown by the solid line. The lower critical field $\mu_0H_{c1}$ was determined from the field where the magnetization deviates from the linear response by a small value of 0.1~emu/cm$^3$, as displayed in the inset of Fig. \ref{Figure4}(c). Figure \ref{Figure4}(c) displays the temperature dependence of the obtained $\mu_0H_{c1}$ after correcting for demagnetization effects, which were fitted with
\begin{equation}\label{equation5}
\mu_0H_{c1}(T) = \mu_0H_{c1}(0)\left[1-\left(\frac{T}{T_c}\right)^2 \right]
\end{equation}
yielding a zero temperature value of $\mu_0H_{c1}(0)$ = 10.1(2)~mT.

Figure \ref{Figure5} displays the temperature dependence of $\rho(T)$ and $C_e/T(T)$ of La$_4$$T$Al under various applied fields. The superconducting transitions are gradually shifted to lower temperatures and become slightly broadened with increasing magnetic field. The specific heat jump becomes smaller, suggesting second-order transitions in field for the three materials, which is characteristic of type-II superconductivity. The upper critical fields $\mu_0H_{c2}$ plotted as a function of temperature are shown in Figs. \ref{Figure5}(g)-(i). Here $T_c$ are determined from where there is zero resistivity (blue) and the midpoints of the specific heat jump (red). Using the slope of the upper critical field near $T_c$ ($d\mu_0H_{c2}/dT$ = $-$1.04~T/K, $-$1.49~T/K, and $-$1.16~T/K for the Ru, Rh, and Ir variants, respectively), the upper critical fields $\mu_0H_{c2}(T)$ can be fitted by the Werthamer-Helfand-Hohenberg (WHH) model \cite{WerthamerPhysRev1966}, with zero-field values $\mu_0H_{c2}(0)$ of 1.29(1)~T, 3.31(2)~T and 1.34(1)~T, respectively. All of these are significantly below the Pauli limit ($\mu_0H_{c2}^{\mathrm{Pauli}}$ = 1.86$T_c$). The Ginzburg-Landau coherence length $\xi_{\mathrm{GL}}$ was calculated to be 15.9(1)~nm for La$_4$RuAl, 9.9(1)~nm for La$_4$RhAl and 15.6(1)~nm for La$_4$IrAl, using
\begin{equation}\label{equation6}
\mu_0H_{c2}(0) = \frac{\Phi_0}{2\pi\xi_{\mathrm{GL}}^2}
\end{equation}
where $\Phi_0 = h/2e$ is the magnetic flux quantum. Using the lower critical value $\mu_0H_{c1}(0)$ = 10.1(2)~mT (determined from magnetization measurements), the superconducting penetration depth $\lambda_{\mathrm{GL}}$ for La$_4$RhAl can be estimated to be 225(2)~nm via
\begin{equation}\label{equation7}
\mu_0H_{c1}(0) = \frac{\Phi_0}{4\pi\lambda_{\mathrm{GL}}^2}\mathrm{ln}\frac{\lambda_{\mathrm{GL}}}{\xi_{\mathrm{GL}}}
\end{equation}
Therefore, the Ginzburg-Landau parameter $\kappa_{\mathrm{GL}}$ = $\lambda_{\mathrm{GL}}/\xi_{\mathrm{GL}}$ is 22.7(3), which is much larger than $1/\sqrt{2}$, indicating type-II superconductivity in La$_4$RhAl. The thermodynamic critical field $\mu_0H_c(0)$ can be estimated from
\begin{equation}\label{equation8}
\mu_0H_{c1}(0)\mu_0H_{c2}(0) = (\mu_0H_{c}(0))^2\mathrm{ln}\kappa_{\mathrm{GL}}
\end{equation}
yielding 104(3)~mT for La$_4$RhAl.

The residual electronic specific heat coefficient $\gamma_0(H)$ can give information about low-energy quasi-particle excitations near vortex cores and hence can characterize the superconducting pairing state. The magnetic field dependence of the normalized residual Sommerfeld coefficient $\gamma_0(H)/\gamma_n$ are shown in the insets of Figs. \ref{Figure5}(g) and (h), where the linear behaviors are consistent with $s$-wave superconductivity.

The upper critical fields of La$_4$RuIn and La$_4$IrIn were determined from $\rho(T)$ and $C_e/T$ in various magnetic fields, as shown in Fig. \ref{Figure6}. The $\mu_0H_{c2}$ values are displayed in Figs. \ref{Figure6}(e) and (f). The WHH model can be used to fit the data yielding $\mu_0H_{c2}(0)$ of 0.22(1)~T and 0.50(1)~T for La$_4$RuIn and La$_4$IrIn, respectively, where both values are much smaller than the Pauli limit. The coherence length $\xi_{GL}$ are calculated via Eq. \ref{equation6} to be 38.7(1)~nm (La$_4$RuIn) and 25.6(1)~nm (La$_4$IrIn). The corresponding data are summarized in Table. \ref{table:table1}.

  \begin{figure}
  	\includegraphics[angle=0,width=0.49\textwidth]{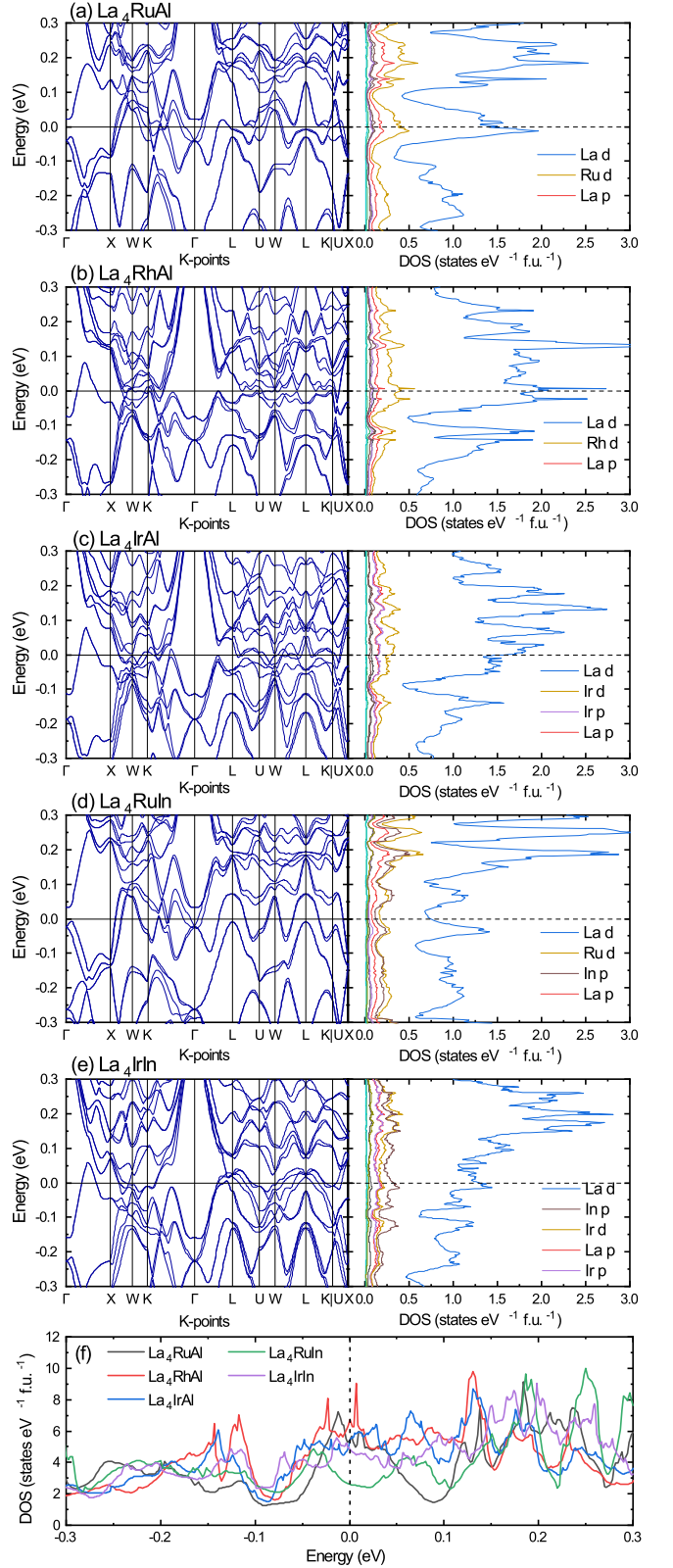}
  	\vspace{-12pt} \caption{\label{Figure7}  DFT calculations of the electronic band structure and partial density of states of (a) La$_4$RuAl, (b) La$_4$RhAl, (c) La$_4$IrAl, (d) La$_4$RuIn, and (e) La$_4$IrIn, plotted in the energy range $\pm0.3$~eV from the Fermi level, taking spin-orbit coupling into account. (f) Total density of states of the five superconductors near the Fermi level. }
  	\vspace{-12pt}
  \end{figure}

\begin{table*}[tb]
\caption{Superconducting and normal state properties of La$_4$$TX$. Here $N_{\mathrm{DFT}}(E_F)$ and $\gamma_n^{cal}$ correspond to the calculated values of the density of states at the Fermi level and the Sommerfeld coefficient, respectively, from DFT calculations.}
\label{table:table2}
\begin{ruledtabular}
 \begin{tabular}{ c c c c c c c c c c c c}
{}      &$T_c$   &$\mu_0H_{c2}(0)$   &$\xi$    &$\gamma_n$     &$\theta_D$   &$\lambda_{\rm el-ph}$    &$N_{\mathrm{DFT}}(E_F)$   &$\gamma_n^{cal}$  &$E_{\mathrm{ASOC}}$  &$\frac{E_{\mathrm{ASOC}}}{k_BT_c}$    \\
{}     &{(K)} &{(T)}  &(nm) &{(mJ~mol$^{-1}$~K$^{-2}$)}  &{(K)}  &{}   &{(states eV$^{-1}$ f.u.$^{-1}$)} &{(mJ~mol$^{-1}$~K$^{-2}$)}  &{(meV)}      &{}                                 \\
\hline\\[-2ex]
{La$_4$RuAl}  &{1.77} &{1.29(1)} &{15.9(1)} &{20.1(1)} &{204(2)} &{0.45$-$0.54} &{5.51} &{20.0} &{13} &{89}   \\
{La$_4$RhAl}  &{3.05} &{3.31(2)} &{9.9(1)}  &{29.3(2)} &{208(2)} &{0.52$-$0.62} &{6.35} &{24.3} &{16} &{63}   \\
{La$_4$IrAl}  &{1.54} &{1.34(1)} &{15.6(1)} &{21.3(1)} &{191(1)} &{0.44$-$0.53} &{5.18} &{18.7} &{37} &{285}   \\
{La$_4$RuIn}  &{0.58} &{0.22(1)} &{38.7(1)} &{15.9(1)} &{169(1)} &{0.36$-$0.45} &{2.53} &{8.7}  &{17} &{341}   \\
{La$_4$IrIn}  &{0.93} &{0.50(1)} &{25.6(1)} &{21.1(2)} &{177(1)} &{0.39$-$0.48} &{4.90} &{17.2} &{40} &{510}   \\
\end{tabular}
\end{ruledtabular}
\end{table*}
\subsection{\uppercase\expandafter{d}. DFT calculations}

Calculations of the band structure and density of states (DOS) of La$_4$$TX$ ($T$ = Ru, Rh, Ir; $X$ = Al, In) were performed, taking into account spin-orbit coupling (SOC), which are shown in Fig. \ref{Figure7}.
Since the crystal structure contains four formula units per unit cell, the band structure is very complex and highly metallic, consistent with resistivity measurements. The electronic bands are clearly split due to the ASOC and the splitting magnitude near the Fermi level $E_{\mathrm{ASOC}}$ for each compound is listed in Table. \ref{table:table1}. The values are between 13 and 40~meV, which are moderately large compared with other NCSs \cite{Smidman2017}. As the SOC is related to the atomic number $Z$, the compounds with the heavier elements Ir and In have larger $E_{\mathrm{ASOC}}$ splitting than those with Ru, Rh, and Al. In addition, overall the Rh/Ir compounds can be regarded as electron doped compared to the Ru compounds. This effect is most prominent in the states around $\Gamma$, where the electron pockets in the Ru compounds are absent in Rh/Ir compounds. It is noted that numerous band-crossing-like features between X-W and U-L exist near the Fermi level in this compound. The band topology and its relation with the superconductivity in this family requires further study.

The partial DOS are shown on the right side of the corresponding band structure figures. The contributions to the DOS at the Fermi level $N(E_F)$ are dominated by La-$d$ orbitals for all the materials, with smaller contributions from T-$d$, La-$p$ and In-$p$ orbitals. In La$_4$RhAl, the quasi-flat bands near the $W$ point at $\sim$ $E_f$$-$30~meV and $\sim$ $E_f$+10 meV result in two van-Hove singularity like peaks in the DOS, which disappear in La$_4$IrIn as the SOC is enhanced. The values of the total $N(E_F)$ are summarized in Table. \ref{table:table1}, where it can be seen that the compounds with relatively higher $T_c$ have comparatively larger $N(E_F)$. The relatively lower $T_c$ values and larger band splitting of La$_4T$In may point to influence from the ASOC on  $N(E_F)$  and $T_c$. The electronic specific heat coefficient can be estimated using
\begin{equation}\label{equation9}
\gamma_n^{cal} = \frac{1}{3}\pi^2k_B^2N(E_F)(1+\lambda_{\mathrm{el-ph}})
\end{equation}
where $\lambda_{\mathrm{el-ph}}$ were experimentally determined as described above. The resulting values $\gamma_n^{cal}$ = 20.0 (La$_4$RuAl), 24.3 (La$_4$RhAl), 18.7 (La$_4$IrAl), 8.7 (La$_4$RuIn), and 17.2 (La$_4$IrIn) mJ~mol$^{-1}$~K$^{-2}$ are slightly smaller than the observed $\gamma_n$ (Table. \ref{table:table1}), suggesting a small effective-mass enhancement due to electronic correlations.

To explore the relationship between superconducting properties and the ASOC strength, the ratio $E_{\mathrm{ASOC}}/k_BT_c$ for the five superconductors are listed in Table. \ref{table:table1}. La$_4$RuAl and La$_4$RhAl have low $E_{\mathrm{ASOC}}/k_BT_c$ values (89 and 63) comparative to those of nodeless Li$_2$Pd$_3$B (44). La$_4$IrAl, La$_4$RuIn and La$_4$IrIn have larger values of 285, 341 and 510, respectively, which are comparable to the nodal superconductors Li$_2$Pt$_3$B and CaPtAs with respective values of 831 \cite{LeePhysRevB2005} and 780 \cite{Xie2020}. As such, it is of particular interest to probe the pairing states of these systems at lower temperatures, to look for unconventional superconducting properties. Furthermore, the calculated partial DOS of La$_4$$TX$ indicate that there is a dominant contribution from La orbitals near $E_F$, and hence substitution on the La-site may allow for a wider range of ASOC splitting to be realized.

\section{\uppercase\expandafter{\romannumeral4}. Summary}

We report superconductivity in a series of noncentrosymmetric compounds La$_4$$TX$ ($T$ = Ru, Rh, Ir; $X$ = Al, In) with the cubic Gd$_4$RhIn-type structure (space group $F\bar{4}3m$), and characterize the superconducting properties by means of electrical resistivity, magnetization and specific heat measurements, as well as electronic structure calculations. Bulk superconductivity was found with $T_c$ values of 1.77~K, 3.05~K, 1.54~K, 0.58~K, and 0.93~K for La$_4$RuAl, La$_4$RhAl, La$_4$IrAl, La$_4$RuIn, and La$_4$IrIn, respectively. Both the temperature and field dependences of the specific heat of La$_4$$T$Al are consistent with $s$-wave superconductivity, but confirmation of the gap structure requires different measurements sensitive to low energy excitations to be performed at lower temperatures. All the systems exhibit moderate upper critical fields, significantly less than the Pauli limit, indicating the dominance of orbital limiting. In the case of La$_4$RhAl, magnetization measurements show typical behavior of type-II superconductivity, with a Ginzburg-Landau parameter of 22.7(3). Electronic structure calculations reveal that the electronic bands near the Fermi level are split due to the ASOC, with a range of magnitudes of the band splitting that is larger in systems with heavier elements. Our results suggest that the  La$_4$$TX$ systems are a large family of noncentrosymmetric superconductors, where the effect of the ASOC on the bands near the Fermi level are highly tunable by elemental substitution, and therefore these are good candidates for examining the relationship between the ASOC and superconducting properties. As such, it is of particular interest to characterize the superconducting pairing states of these compounds, via both measurements to lower temperatures which are sensitive to the superconducting gap structure, as well as probing for broken time reversal symmetry using muon-spin relaxation or the Kerr effect.

\section{Acknowledgments}

This work was supported by the Key R\&D Program of Zhejiang Province, China (2021C01002), the National Natural Science Foundation of China (No. 11874320, No. 12034017, No. 11874137, and No. 11974306), the National Key R\&D Program of China (No. 2017YFA0303100), and Zhejiang Provincial Natural Science Foundation of China (R22A0410240).

\bibliographystyle{apsrev4-1}

\end{document}